\documentclass[12pt]{article}
\usepackage{amssymb}
\usepackage{epsfig,color}
\usepackage{pstricks,graphicx,epsfig,color,amssymb,amsmath,amscd}
\usepackage{cite}
\usepackage[]{graphicx}
\usepackage{makeidx}
\usepackage{amsmath}
\usepackage{nccmath}
\usepackage{setspace}

\usepackage[]{caption}
\renewcommand\rho{\varrho}

\newcommand{\be}{\begin{eqnarray}}
	\newcommand{\ee}{\end{eqnarray}}
\newcommand{\rar}{\rightarrow}

\newcommand{\tcblue}{\textcolor{blue}}

\newcommand{\tcmag}{\textcolor{magenta}}

\newcommand{\tcviol}{\textcolor{violet}}

\title{\textbf{Progress in Particle Physics and Modern Cosmology}}
\author{A. D. Dolgov}
\date{}

\begin{document}

\maketitle

This is the translation into English of my paper published in "Einshteinovskij Sbornik"  \\
1980-1981~\cite{dolgov-fen}.
The content of the book is presented in the Appendix.
This is the first paper where the dynamical mechanism of "phoenix universe" was worked out.
The notion of phoenix universe was first mentioned in the paper by Lemaitre in 1933~\cite{lemaitre},
where he assumed that a repetition of successive phases of expansion and contraction was possible. 
Lemaitre called such model of the universe, that is born, dies and is reborn, the phoenix model, 
named after the mythical bird able to reborn from the ashes. According to the author,
there could be an infinite number of such cycles in the past and future.
In this model, however, a possible solution to the fundamental 
question was unclear, what was the  mechanism for the  “rebirth” of the universe? This mechanism was
worked out in ref.~\cite{dolgov-fen}, published in 1980 in Russian. Moreover, in our recent 
paper~\cite{eva-dad} a mechanism of dynamical cancellation of vacuum energy was proposed,
that permits to eliminate vacuum energy locally down to zero and permits universe to jump 
to a lower hot level, leading to rebirth of a hot universe.

\section*{Introduction}

Modern cosmology was born as a result of Einstein's formulation~\cite{Ein} of the general theory of relativity 
and Friedmann's discovery\cite{Fried} of non-stationary solutions of Einstein's equations. 
Einstein, when he first expressed the idea of applying the equations he had discovered to cosmology was, however,
discouraged by the fact that these equations do not have stationary solutions in a cosmological situation.
To eliminate this "deficiency," Einstein proposed generalizing the equations of general relativity by adding the so-called cosmological 
term~\cite{Ein-Lam}, which could 
stabilize the Universe. However, it soon became clear that the Universe is, after all, expanding in full accordance with Friedmann's predictions. This 
discovery was made by Hubble~\cite{Hubble}, who saw that distant astronomical objects are moving away from us at a 
speed proportional to distance to them. The 
natural next step was the formulation of the hot model of the Universe by Gamow~\cite{Gamow}, 
based on the theory of synthesis of elements, 
which became the generally accepted cosmological model 
after the discovery of the relic electromagnetic radiation by Penzias and Wilson~\cite{Penz-Wil}.
The status of the hot model was further strengthened after detailed calculations of abundances of 
light elements produced in 
a hot Universe that were performed by Wagoner, Fowler, and Hoyle~\cite{WFH}. 
The results of these calculations were in excellent agreement with astronomical observations. Particularly successful was the coincidence with observations of the calculated $^4 He$ abundance, which had not been achieved in other models.

The impressive achievements of Friedmann's cosmology further highlight the fact that the
assumptions underlying it are truly mysterious. Of course, we are not talking about the theoretical basement
which is very simple and beautiful; the question concerns the choice of model parameters, the
initial conditions that determine the development of the universe. It is precisely the realization
of the maximally symmetrical, homogeneous, isotropic initial state, very similar in its properties to vacuum.
This requires a precise matching of the model parameters with precision, which has no analogues
in physics. 

Otherwise, the world would be completely different, unsuitable for life, at least in its current
form. One might think that the Creator took special care to prepare comfortable conditions for
us, taking care of  each one from $10^{80}$ particle of the visible part of the world to make it suitable for our life. 

This circumstance is the basis for the so-called anthropic principle in its strong formulation: the fact 
of our existence is the answer to the question of why the universe is the way it is, and why
we are in the universe, since in the universe not adapted for life such a question cannot be asked, 
because  there  simply would be no one to ask it. 
In such a situation, physicists have nothing to do. I would therefore like
to find some kind of explanation for these "fundamental" problems of cosmology, to construct a
model in which the universe developed to its current state more or less independently of the
initial conditions, adhering to the fundamental laws of physics. Later, thanks to revolutionary
advances in elementary particle theory, it became possible to do this. In this sense, cosmology
is entering a new level: fundamental cosmological parameters, considered as given values, as
a result of the choice of initial conditions, may turn out to be calculable quantities. However, it
should not be assumed that all these cosmological problems have already been solved; there
are still many difficulties ahead, and, possibly, the final answer will differ significantly from the
variants currently under consideration, but in any case, there is a fundamental possibility of
answering the question about the selection of initial conditions in the "best of all possible
worlds" scenario.

The order of presentation of the material in the article is the following: in section 1, the basic
observational facts  about the universe, 
that will be relevant  for the future content, are  briefly discussed .
Section 2 examines important cosmological problems and indicates possible ways of solving
them within the framework of the so-called model of an expanding universe, the inflationary
model. Section 3 provides a more detailed description of the inflationary model, its difficulties,
and possible ways to overcome them. The conclusion summarizes the results.

\section*{\bf Universe today (observational data)}

1. The fact of the expansion of the universe is not disputed by
anyone. Apparently, it does not cause any objections to the Hubble's law 
of the proportionality of the speed of an object to the distance of
that object
\be
v = H r
\label{hR}
\ee
this is true of course on the average, with exclusion of the chaotic
motion of individual galaxies in their clusters. 
So far, however, there is no consensus on the value of the coefficient of proportionality, known as
the Hubble constant $H$.
Most astronomers currently give values of $H$ close to 100km/sec/Mps\cite{vauc}
but there are works~\cite{sandage}  in which a twice smaller value is presented.
The criticism of the latter paper by the  supporters of the large value of H seems convincing but 
the data on  the universe age (see section 6 below) force us to
conclude that  $H= 100$ km/sec/Mpc remains a viable possibility.

2. Whether the expansion will stop or continue indefinitely
is determined by the ratio of the average energy density in the
universe to the so-called critical density~\begin{footnote}{We use here the system of 
units used in particle physics, where speed of light, 
Boltzmann constant, and reduced Planck constant 
are all equal to unity: $C = k = \bar h = 1 $.
For example the proton mass is equal to 
$m_p = 940 MeV = 10^{13} K = 5\cdot 10^{13} cm^{-1}= 1.5\cdot 10^{24} sec^{-1}$. }
\end{footnote} 

\begin{equation}
\Omega = \rho/\rho_c, \,\, \rho_c = 3H^2/(8 \pi G) = 1.86 \cdot 10^{-29} h_{100}^2 \,\,{\rm g/cm}^3
\label{Omega}
\end{equation}
where $G$ is the gravitational constant: $G \equiv m_P^{-2} = \left(1.22\cdot 10^{19} {\rm GeV} \right)^{-1}$ and
\begin{equation}
h_{100} = H /( 100\, {\rm km/sec/Mpc} ) . 
\label{h-100}
\end{equation}

If $\Omega  > 1$, then the universe is closed and a period of contraction will eventually take place; 
if $\Omega \leq 1$ , then expansion will continue indefinitely. 
The case $\Omega = 1$ corresponds to the  spatially flat, Euclidean universe, 

Modern estimates of the average energy density~\cite{gott}, based on
measurements of its gravitational effects, are close to $\Omega =0.3$, which
supports the open universe model. 

It is interesting that  the determination of the matter density by direct estimate of the amount of matter contained in
visible objects and interstellar space, gives the result  approximately an order of magnitude smaller,
$\Omega_B \sim 0.03$.  The sub-index $B$ indicates that this quantity is  the usual proton-neutron  or baryonic matter. The discrepancy between $\Omega$ found from dynamics of galaxies and $\Omega_B$ creates the 
problem of dark matter of the Universe~\cite{einasto}.\footnote{ 
Fritz Zwicky who discovered dark matter in the 30th and was badly criticized by the community.
referred contemptuously to  “the useless trash in the bulging astronomical journals”, saying 
{“Astronomers are spherical bastards. 
No matter how you look at them they are just bastards.”}}

One might think that  by some not yet known reasons we do not
see a significant part of the usual proton-neutron-electron matter. 
However, given the abundance of deuterium and helium-4
in the Universe, on the one hand, and the theory of galaxy formation, on the
other, this possibility is unlikely. The most popular view is that the invisible matter is either
massive neutrinos or some kind of weakly interacting particles, such as axions or photinos.
A discussion of these issues in the literature can be found in ref.~\cite{zeld}.  A modifications of
gravitational interaction cannot be ruled out as well.

In principle, an invisible matter could lift $\Omega$ up to unity but only  if it is distributed homogeneously
throughout all the space\cite{rees}, though is seems  quite unlikely, but strictly speaking not excluded, 
This may be of interest for the model of inflationary universe cosmology discussed below.

3. As is well known, the General Relativity equations allow for a generalization
by introducing the so-called cosmological term, $\Lambda$~\cite{Ein-Lam}:
\be
R_{\mu\nu} - (1/2) g_{\mu\nu} R = 8 \pi G T_{\mu\nu} - \Lambda g_{\mu\nu},
\label{cosm-term}
\ee
where $T_{\mu\nu}$ is the energy-momentum tensor of matter, and the term proportional to $\Lambda$ describes 
gravitating vacuum and can be written in the form with  $\rho_{vac}$ being the vacuum energy density:
\begin{equation}
\Lambda g_{\mu\nu} = -8\pi G T_{\mu\nu}^{vac}=-8\pi G \rho_{vac} g_{\mu\nu} .
\label{lambda-value}
\end{equation}
 
In the standard scenario
of the universe evolution it is assumed that $\Lambda = 0$, though the observational bounds evaluated in 
units of the critical energy density are not very strong:
\begin{equation}
| \rho_{vac} | <  5\cdot 10^{-47} m_N^4 \approx 10^{-29} g/cm^3  \sim \rho_c .
\label{rho-vac-lim}
\end{equation}

On the other hand, the vacuum energy is very small compared to the characteristic values of $\rho$ at the time of the
universe creation and in view of that it looks natural to assume that $\rho_{vac}$ identically vanishes.
We will return to this issue below.

4.  The averaged over  large cosmological scales distribution
of matter in the universe is highly uniform. Of course, the inhomogeneities are huge at the galactic scales
but at distances greater than about 100 Mpc, with the variations of density over directions are quite small
\begin{equation}
\Delta \rho /\rho < 10^{-3} .
\label{delta-rho}
\end{equation}

The homogeneity and isotropy of the universe is also supported by the observations of cosmic microwave 
background radiation which directional vaariations  do not exceed  $10^{-4}$

5. There are compelling observational reasons to believe that
antimatter is practically absent in the universe, i.e. positrons,
antiprotons, antineutrons, but it follows from theory that the amount of
antimatter is most likely not insignificant. Strictly speaking, it cannot
be ruled out that distant galaxies consist of antimatter, but in all
known cases of colliding galaxies or galaxies surrounded by common
clouds of interstellar gas, it is clear that these regions contain matter
of the same type. This circumstance, as well as the fact that there are few
antiprotons in cosmic rays, makes the hypothesis of the existence of
antimatter in significant amount highly unlikely. The presence
only of matter in the surrounding universe is called the charge or baryon asymmetry
of the universe.

An important constant in cosmology is the ratio of the average density of baryons $N_B$
to the density of relic photons $N_\gamma$:
\begin{equation}
N_\gamma = 550 (T/3{\rm K})^3 {\rm cm}^{-3} .
\end{equation}
According to current data, this ratio is in the range
\begin{equation}
\beta = N_B/N_\gamma = 10^{-9} - 10^{-10}.
\end{equation}

6. The time elapsed from the hot singularity to the present time,
is called the age of the Universe $t_U$. The value of $t_U$ is surely  
greater than the estimated age of the Earth:
$5\cdot 10^9$ years. Nuclear chronology, as well as the theory of stellar evolution, together with the fact of the
observations of old star clusters, leads to a much larger value~\cite{t-u} 
\begin{equation}
t_u \approx15 \cdot 10^9\,{\rm years} .
\end{equation}

Theory permits to express the age of the universe through 
the current values of Hubble's constant and parameter $\Omega$. Since, according to the
standard scenario, the universe spent most of its life in a state of dominance of non-relativistic matter the following 
expression is valid:
\begin{equation}
t_u = 10.8 \cdot 10^9\, {\rm years} \,[ h_{100} ( 1+ \sqrt{\Omega}/2 )]^{-1}
\end{equation}
(under assumption that the cosmological constant vanishes).

So we have to choose between the following possibilities:\\
a)  the Hubble parameter is smaller than that presented in majority of papers,  $h_{100} < 0.6$;\\
b)  nuclear chronometry and stellar evolution theory suggest a significantly larger value: 
$t_u \approx 15 \cdot 10^9$ years;\\
c) the cosmological constant is non-zero and close to its upper limit on $\rho_{vac} $ (\ref{rho-vac-lim}),
permitted by astronomical observations.

Let us note, anticipating what is written below, that in the inflationary universe model these contradictions become even
more profound, since in this model the value $\Omega = 1$ is predicted and the universe age $t_u$ should be 
smaller than that in the case $\Omega = 0.3$.

\section*{Fundamental Cosmological Problems.}

The Einstein equations which make the basis of the modern cosmology have the following very simple form for 
homogeneous and isotropic distribution of matter:
\begin{eqnarray}
\ddot  a &=& \frac{4\pi G}{3}\,a \left(3 p + \rho \right), \label{ddot-a} \\
\dot a^2 &=& \frac{8\pi G}{3}\,\rho a^2 - k  \label{dot-a}.
\end{eqnarray}
where dots denote differentiation over time, $\rho$ and $P$ are respectively energy density and pressure of matter
(with possible inclusion of the vacuum term); $a$ is the scale factor, the value of
which is not determined. If $k \neq 0$, the value of $a$ can be normalized  by the conditions $k=+1$ or $k = -1$.

Equation (\ref{dot-a}) can be conveniently rewritten as
\begin{equation}
\rho = \rho_c - \frac {k}{a^2},\,\,\, \rho_c = \frac{\dot a^2}{8\pi G a^2},
\label{rho-of-a}
\end{equation}
from which it follows that $k<0$ corresponds to a closed 
universe, and $k > 0$ corresponds to an open one.

There are different expansion regimes depending on the equation of state. Assuming $k =0$ we find the following. \\
Relativistic gas: $\rho \sim a^{-4}, a \sim t^{1/2}$:
\begin{equation}
p =\rho/3, \,\,\, \rho(a) \sim a^{-4}, \,\,\,\,\, a \sim t^{1/2}; 
\label{rel-regime}
\end{equation}
Non-relativistic gas: $\rho \sim a^{-3}, a \sim t^{2/3}$:
\begin{equation}
p =0, \,\,\, \rho(a) \sim a^{-3}, \,\,\, a \sim t^{2/3};
\label{nonrel-regime}
\end{equation}
Cosmic strings:
\begin{equation}
p =-\rho/3, \,\,\, \rho(a) \sim a^{-2}, \,\,\, a \sim t;
\label{strings}
\end{equation}
Domain walls:
\begin{equation}
p =-2\rho/3, \,\,\, \rho(a) \sim a^{-1}, \,\,\, a \sim t^2;
\label{walls}
\end{equation}
Gravitating vacuum:
\begin{equation}
p =-\rho, \,\,\, \rho(a) = const, \,\,\, a \sim  \exp\left[ \sqrt{\frac{8\pi G \rho}{3}}\,t  \right] .
\label{vacuum}
\end{equation}

At the present time the universe is dominated by nonrelativistic matter and the expansion regime is close to
(\ref{nonrel-regime}), if $\rho$ is not too much different from $\rho_c$.
The transition from relativistic regime to  non-relativistic one takes place at the redshift $ z = 4\cdot 10^{4} h_{100}^2$.
It is unknown if there were regimes  dominated by cosmic strings or domain walls, but it is often assumed that there was
period of cosmolgical term dominance,

The cosmological impact of domain walls which appears at spontaneous discrete symmetry breaking was considered 
in ref.~\cite{walls}, where it was pointed out that these walls, if they existed, would destroy the homogeneity of the universe. 

The role of cosmic strings was studied in papers~\cite{strings}, where it was shown that the inhomogeneities created by
strings are safely small and, moreover, strings that appear in unified theories of strong and electroweak interactions
with characteristic energy scale $10^{14} - 10^{15} $ GeV could explain the observed 
the large scale structure of the universe 
in the forms of galaxies or their clusters.

How far can we travel backward in time depends upon our knowledge  of the particle interactions at high energies and
densities. One may be sure  that the simple relativistic  expansion regime  (\ref{rel-regime}) was realized up to energies 
of several tens or even hundreds MeV. Somewhere close to these energies  the equation of state of the primeval plasma
might be changed because of  the QCD phase transition from free quarks and gluons down to hadrons. Prior to this phase 
transition the equation of state of the primeval plasma is also close to 
that of the ideal gas. According to our present day 
knowledge, confirmed by  laboratory experiments, this is possibly true starting down from the temperatures about
100 GeV.  Advancing into the region of higher temperatures is not so reliable, since experimental data on the particle 
properties at $E >100$ GeV are practically absent.

On the other hand, there is a theory that describes strong
and electroweak interactions of elementary particles in a unified way and has a number of
other attractive features, which asserts that nothing revolutionary happens up to the Planck
energy $E \approx 10^{19}$ GeV when, apparently, the effects  of quantum gravity become significant.
It is not yet clear how to deal with this. 

Within the framework of such theories, it can be concluded 
that the dynamics of the universe is in principle known up energies $E_{PL} = 10^{19} $ GeV.
If we assume that all phase
transitions occurring during the cooling of the Universe are phase transitions of 
of the second type or
weakly delayed transitions of type I, then the influence of these transitions on the nature
of expansion will not be particularly noticeable and the expansion regime (\ref{rel-regime})
would be approximately valid up to the Planck energy.

The statement about phase transition in theories with spontaneously broken symmetry was  first made 
in ref.~\cite{phase-tran} and since then it has been used in cosmological models.

Returning to equation (12), we rewrite it in the form:
\begin{equation}
\Omega_2^{-1}  - 1 = \left( \Omega_1^{-1}  - 1\right) \, \frac{\rho_1 a_1^2}{\rho_2 a_2^2} .
\label{Omega-of-a}
\end{equation}
where the indices 1 and 2 refer to the values of the quantities at times $t_1$ and $t_2$. 
Taking $t_1$ as the present moment and $t_2$ as the moment of transition
from relativistic to non-relativistic expansion law we find:
\begin{equation}
\Omega_2^{-1}  - 1 = \left( \Omega_1^{-1}  - 1\right) \, z_2^{-1} \approx 10^{-4}.
\end{equation}
Choosing $t_2$ as the beginning of the primordial nucleosynthesis: $t_2 = 1$ sec and $T_2 = 1$ MeV,
 and in this state the universe surely was, as is proven by the abundance of light elements, 
we find $\Omega_2^{-1}-1 \approx 10^{-16} $. If we move further deeper to the beginning  up to
$ T_2 = T_{Pl} = 10^{19} $ GeV, we find that  $|\Omega_2 -1 | = 10^{-59}$.  In other words,  for the Universe to reach
the present day state the initial state should be fiine-tuned with fantastic accuracy  $|\Omega_2 - 1| \approx   10^{-59}$.
The result looks natural enough since for  $|\Omega_2 -1 |\sim 1$, the characteristic time 
when the universe expansion 
for a closed universe would turn into contraction during time close to the Planck one, while an open universe 
would expand so fast that neither galaxies nor stars or planets would be formed. Thus our existence demonstrates the
extremely fine-tuned initial state of the universe close to the flat  3D one ($k=0$).

This mysterious fact is the subject of one of the most pressing problems
in cosmology: how did such favorable extremely fined tuned initial conditions could arise? This
problem is called {\bf the problem of initial conditions} (the first problem).

There are several other cosmological problems, without solving which we cannot be sure
that we understand how our world was created.\\
{\bf The second problem:  Isotropy and homogeneity} of the universe also imply very specific
initial conditions, noticeable deviations from which are not allowed even by the anthropic principle.
This makes it all the more desirable to find a natural explanation for this fact.\\
{\bf The isotropy of the cosmic background radiation} presents a new problem (the third), 
which is the so called the horizon problem. It is related to
the fact that the scale factor  in the regimes  (\ref{rel-regime},\ref{nonrel-regime}) rises 
slower than the size of the causally connected region $r_h \approx t$.

Relic radiation became free, i.e., it ceased to interact with anything at the hydrogen
recombination temperature $T_{rec} \approx 3000$ K. This corresponds to the change of the scale factor
by about the factor of 1000. For the standard model of the universe evolution this corresponds to the horizon size 
at this moment equal to $\sim 10^{13} $ sec. Correspondingly the size  of the region of the physical processes which
could create background radiation cannot exceed  this value.
 
The attempt to explain this phenomenon, based on the hypothesis of the existence of a
maximum density of energy, was made in~\cite{entropy}, but the question of the existence of a maximum
density of energy was not resolved. An attempt of this kind, based on the hypothesis of the
existence of maximum energy density, was made in ref.~\cite{entropy}, but the question of entropy growth in
this model remains unclear.

{\bf The problem of charge asymmetry in the universe} (i.e. the presence in the universe only of
particles (i.e., protons, neutrinos, electrons) and practically  absence of antiparticles, which
has been a problem for cosmologists for many years, now can be considered as solved according to
ref.~\cite{sakharov}.

Within the frameworks of the model, there is not only qualitative but also quantitative agreement with
astronomical data for the value of $N_b/N_\gamma$. To avoid returning to this question later, let us
briefly review main features of  the mechanism of generation of  the excess of particles 
over antiparticles in the universe. A more detailed discussion can be found in
review~\cite{BG-rev}  and in the popular paper~\cite{AD-YaZ}.

The basic hypothesis is that the baryon charge B is not conserved.
Such conservation is indeed predicted by models of grand unification. In these models,
there are superheavy particles with masses of the order of $10^{14}-10^{15}$ GeV
decaying into the states with different values of the baryonic number. 
Two other essential ingredients of the model are the
violation of charge invariance (i.e., the difference in interactions between particles and
antiparticles) and deviation from thermodynamic equilibrium in the expanding
universe.  It can be shown that under such conditions, the decay of the aforementioned
superheavy particles (X-, or H-bosons) wiill lead to an excess of particles over antiparticles 
which because of the violation of thermal equilibrium  will not be compensated for by other
processes.  Although the magnitude of this excess cannot be precisely calculated, since the details
of the interaction between X-f-bosons are unknown, order-of-magnitude estimates give an
answer that is reasonable and consistent with observations. For a number of the models the result 
does not depend on the initial conditions,
i.e. on whether there was initiallyy an excess of baryons or antibaryons, or whether the plasma was
charge neutral. An essential feature of this scheme in its classical version
is the presence of a large number of X- or Z-bosons in the initial state of the plasma, 
for which it is necessary to reach the temperatures about $T\approx 10^{15}$ GeV.

{\bf The seventh problem, the problem of magnetic monopoles,} 
is somewhat different from those listed above, since it is not 
a specifically cosmological problem, but is related to the prediction of the existence of magnetic
monopoles in models of grand unification. Born in the phase transition from symmetric to
asymmetric states during the cooling of the universe, these monopoles would survive until the
present day, and their current concentration, calculated within the framework of the
standard model, turns out to be unacceptably large~\cite{monopoles}.

Problems 1-3 and 7 are uniquely and beautifully solved by the  {\bf inflationary universe model~\cite{guth,linde}},
but at the same time, problem 4 only gets worse. The main idea of the inflationary universe 
model is that at some stage the energy-momentum
 spectrum was dominated by a vacuum energy term, $T^{(vac)}_{\mu\nu} = \rho_{vac} g_{\mu\nu}$. 
According to eq.~(\ref{vacuum}) the scale factor rises exponentially  and the energy density quickly approaches
the critical one, i.e. $\Omega \rightarrow 1$, see eq.~(\ref{Omega-of-a}).

In this model, the universe at a given time moment looks as an expanding
empty space (density of the usual matter exponentially tends to zero). At the same time, all initial
conditions are "forgotten" and the world becomes as uniform as emptiness can be. 
Then the vacuum "matter" explodes, giving birth to elementary particles~\cite{part-prod},
which are thermalized and and the regime of expansion turns into the Friedman one.
The required duration of the exponential (De Sitter) period  $\tau$ depends on the temperature of the created particles.
To ensure $\Omega \sim 1$ at the present time it is necessary that $\exp (H\tau ) > 10^{30} (T/M_{Pl})$, that is
\begin{equation}
H \tau > 70 - \ln (M_{pl} /T)
\label{H-tau}
\end{equation}
This does not look unreasonably large.

If the de Sitter stage actually took place, then at the "zero" moment, parameter $\Omega$ could
have had practically any value. If the universe is open and $\Omega \leq 1$, there no problem arose
as the density of matter decreases, the vacuum energy $\rho_{vac}$ begins to dominate, which does not change in the
course of expansion. If however, $\Omega > 1$ the universe might start to contract when still $\rho_m > \rho_{vac}$
and the exponential stage would not be there.  However, (this comment belongs to L.B. Okun) it is known 
that in the oscillating universe the amplitude
of its oscillations should increase
due to rise of entropy and respectively $\rho_m$ would be
diminished at the the point of maximum expansion. Hence sooner or later a closed universe should come to the
exponential expansion, of course, if $\rho_{vac} > 0$.

Thus, the problem of proximity  of $\rho$ to $\rho_c$ at the present time can be solved 
without hypothesis of  the fine tuning  of the  "initial" parameters. However, if we take such point of view,
then for the explanation of the proximity $\rho$ to $\rho_c$, inflationary scenario is unnecessary since
the swinging universe  will gradually come to then observed present day state
and we just happened to be there when the conditions for life became suitable.

The problem of the horizon in inflationary universe is also naturally solved because the scale factor
$ a \sim \exp {Ht}$ grows faster than horizon.

Note that inflationary model, parameter $\Omega$ should be extremely close to unity because
even a  very small excess of the exponential period duration over its necessary value (\ref{H-tau})
and deviation of $\Omega$ from unity is actually determined by the fluctuations of density in this model.
Astronomical data, however, rather contradicts this, giving a value of $\Omega$ close to 0.3,
unless the universe is filled with homogeneously distributed massive particles, as 
for example, neutrinos~\cite{nu-hom}.

In this regard, it is very important to clarify the value of the Hubble's parameter, 
since for  $\Omega =1$: $h_{100} = 7.2\times 10^9 \,{\rm years} / t_u$.
This may be a realistic way to test inflationary models.

The problem of monopole can be also solved  in this scenario if, after the
phase transition which leads to monopole creation, there is still a noticeable exponential expansion. 
Note that this is required to solve the problem of homogeneity (see above). 
In this case, there would be no more than one monopole~\cite{monopole} in the visible universe, and a
discovery of a second one would require a strong modification of  the standard model. 

Let us note right away that in supersymmetric inflation models~\cite{suzy-infl} 
the problem of the universe revives, and there are possibilities when the number of
monopoles in the universe is non-vanishing but still does not contradict observations 

Return now to the problem of homogeneity. In the first proposed
model~\cite{guth} it was assumed that the exponential expansion
occurred in a symmetric state before transitioning to a non-symmetric
phase. Subsequently, the bubbles of the new phase were formed that filled al the space. 
Generally in such a model the inhomogeneities created by the bubble walls  should be very
large (in ref.~\cite{Ber-Kuz-Tkach} a mechanism of vacuum burning is discussed for which it is
probably not so). Much smaller inhomogeneities arise in the new inflationary scenario~\cite{linde},
according to which considerable exponential expansion took place not only before but also after phase
transition, since in such models the vacuum average of the scalar field (order parameter)  very slowly
in comparison with the universe expansion rate tended to its limiting value.

The problem of inhomogeneities was studied in a series of papers~\cite{inhom}, where it was shown that
in the standard approach based on Coleman-Weinberg $SU(5)$ model (see below) the inhomogeneities would 
be acceptably small only for a very unnatural choice of model parameters. This forced us to turn to supersymmetric 
theories~\cite{susy},  in which such a disadvantage could be overcome. However, in the
supersymmetric approach, other difficulties may arise, related to magnetic monopoles or to particle
production, and generation of the baryon asymmetry, which should be treated with more complicated models

Summarizing, it can be said that, in principle, the inflationary scenario offers a beautiful way to solve a
number of cosmological problems, but there is hardly a natural concrete model that is completely
free from all shortcomings.

Concluding this section, I would like to note that, without solving the problems of
cosmological inflationary models, in which exponential growth is caused by the 
nonzero vacuum energy, 
at least psychologically, does not seem  entirely satisfactory, since it is one thing to assume that
$\rho_{vac} $ is always zero, while it is completely different to assume that at the "beginning" there
was $\rho_{vac} = -\delta \rho $ where $\delta \rho$ is the change of vacuum energy at the phase
transition\footnote{Recently A.D. Linde suggested a model of exponential expansion that occurred without 
strong first order phase transition. It is discussed in more detail in the next section.}.
 (However, defenders of these models may reasonably argue that the cosmological
vacuum energy problem exists independently of inflation.)

An exception is presented by the model suggested in ref.~\cite{starob}, where 
inflation occurs in the pre-Planck era and  is caused by non-linear quantum corrections 
from vacuum polarization to the Einstein equations.
It is also possible that the inflationary model is not the only way to solve the cosmological problems 
under discussion; in fact, there is an alternative attempt~\cite{many-pt}  
based on the hypothesis of a large number of 
phase transitions in the early universe with a huge increase of entropy. 

However, one way or another, inflationary model is the first
cosmological model in which a natural solution of several "eternal" cosmological problems have been
is realized which were previously considered as peculiarities of specific initial
conditions.  In more details, with some technicalities  and analysis of the
difficulties encountered, these issues are discussed in the following section.

Returning to the problem of cosmological constant, we note that, as is
mentioned above, that the data on $t_u$, $H$, and $\Omega$ strongly indicate that $\rho_{vac} \neq 0$.
If inflationary model is valid, and $\Omega = 1$ then for $h_{100} = 1$,  and $t_u = 15\cdot 10^9$ years 
$\rho_{vac}$ should be positive and contribute today approximately $0.95 \rho_c$. 

Since $\rho_c$ varies with time (roughly speaking, as $m_{Pl}^2/ t^2$), and $\rho_{vac}.=const$ it means
 that only at the present stage is the role of $\rho_{vac}$ is noticeable, while at 
 earlier it could be neglected,  which is also one of the mysterious coincidences.

Nowadays there is no  satisfactory model explaining the smallness of the cosmological
constant, but if the presented above values of $H$ and $t_u$ are confirmed, it would be natural to demand 
not a complete compensation of $\rho_{vac}$ but only down to the terms of order $m_{Pl}^2/t^2$.

This, in turn, means that the non-compensated part of $\rho_{vac}$, which,
strictly speaking, is not proportional to $g_{\mu\nu}$ could be noticeable during all the history of the universe 
and to make an impact on the big bang nucleosynthesis, galaxy formation, and to give a contbution to the
hidden mass of the universe. In ref.~\cite{dad-hidden-mass}. a model is suggested  in which vacuum energy
is cancelled down by the condensate of a scalar field  with non-minimal coupling to gravity. The magnitude 
of the condensate increases by the impact of the cosmological term and this negative back reaction 
effectively eliminates  vacuum energy.
The vacuum energy decreases rather slowly, with the remaining value always being
of the order of $\rho_c (t)0$. 

This is achieved due to the very strict and unnatural requirements
imposed by the quantum field theory used in the model.
It can be said that the problem of cosmological constant is the central
problem of cosmology, without whose solution no cosmological model can be
considered satisfactory.

As for the problems of singularity and the "creation of the world," there are
several "crazy" ideas on this subject in the literature. I would like to add one
more to them. Suppose there is, say, a scalar field $\phi$ with an effective
potential that has an infinite number of local minima, 
separated from each other by potential barriers, with each minimum becoming deeper and deeper
as $\phi$ increases. As simplest examples of such potentials one can take
$m^2 \phi^2 \cos (\phi/\sigma) $ or $m \phi^3 [1 - \epsilon\cos(\phi/\sigma)]$.
The universe once created, maybe infinitely long ago, stuck in one of those local minima 
for some time, generally quite long. Then after quantum tunneling this universe or a part of it
it undergoes into another vacuum state with lower energy an so on, so fourth infinitely many times.

The energy released during this phase transition
ends up in the form of elementary particles, which, as a result of the expansion of the universe,
especially in exponential period, "dissolves"  in the universe. 
The cosmological constant (or, rather, cosmological constants) inherent in this model 
can be compensated for by a mechanism of the type described in ref.~\cite{dad-hidden-mass}.

Such a model describes the eternally expanding universe,  infinitely many times 
cooling down practically to a state of vacuum and exploding again. A
suitable name for such a universe would be "Phoenix-Universe." Unfortunately, or perhaps
fortunately, this concept cannot be verified, since in the event of another Big Bang, a possible  observer
will disappear before noticing anything.

\section*{Models of inflating  universe.}
Most models of inflating universe are based on the assumption
are based on the assumption that the phase transition 
from a symmetric to an asymmetric phase is a strongly delayed first order phase
transition. In the symmetric phase the scalar field condensate (order parameter)
is absent, $\langle \phi \rangle = 0$, while vacuum energy is non-zero:
\begin{equation}
T_{\mu\nu}^{vac}  =g_{\mu\nu} V(\sigma),
\label{average-T}
\end{equation}
where $V$ is the effective potential of the scalar field and $\sigma$ is its vacuum average value 
after the end of the phase  transition.
Assumption (\ref{average-T}) does not have any natural physical basis but it is only a demand that cosmological
term disappeared after the phase transition in agreement with observations.

Initially in the universe there could be some matter with energy density $\rho_m$ but  $\rho_m \rar 0$, while 
$\rho_{vac} = const$. If $\rho_{vac}$ would be larger than $\rho_m$ earlier than the phase transition took place,
then the universe would exponentially expand for a while:
\begin{equation}
a \sim \exp( Ht ),\,\,\,\, H = \left[ \frac{8\pi}{3 m_{Pl}^2 \, V(\sigma}   \right]
\label{exp-expand}
\end{equation}
Typical value of $V(\sigma)$  in Grand Unification Models is about $(10^{15} \rm{GeV} )^4$, and hence 
$H \approx 10^{11}$ GeV.

For description of the phase transition let us consider a simple scalar field model with the Lagrangian
\begin{equation} 
L =\frac{1}{2}\left(\partial_\mu \phi \right)^2 - \frac{1}{2}  m_0^2 \phi^2 - \frac{\lambda}{4} \phi^4 + ...
\label{L-of-phi}
\end{equation}
where multidots stand for the terms describing interactions with other fields: gauge bosons, fermions, and
other scalar fields. This Lagrangian is symmetric with respect to the transformation $\phi \rightarrow -\phi $
and possibly some other higher symmetry, if e.g $\phi$ is a milti-component field. In the standard scheme 
of spontaneous symmetry violation it is assumed that the mass squared of the scalar field is negative,
$ m_0^2 <0$ and hence the point  of the stable potential extremum is $ \phi_0^2 = - m_0^2 >0$.
At non-zero temperature the potential acquires an additional term $\alpha \phi^2 T^2$ which could shift 
the equilibrium point to $\phi = 0$~\cite{temp-corr}. So it is clear how field $\phi$ would behave in the course of
the universe cooling down it this simplest model. At hight temperature the average value of $\phi$ vanishes,
$\langle \phi \rangle = 0$. This corresponds to the state with unbroken symmetry. With decreasing temperature 
the sum $m_0^2 + \alpha T^2$ becomes negative and $\langle \phi^2 \rangle = - (m_0^2 + \alpha T^2 )\neq 0 $
and the particles interacting with $\phi$ acquire non-zero masses proportional to $ \langle \phi \rangle$.
It easy to see that the phase transition in this model is the second order 
phase transition, which is not what we need.
However, in more complex models, for example, those based on the
$SU (5)$ group, quantum corrections could lead to the first order phase transition.
For the details and refences to the original literature one can address 
reviews~\cite{pt-first-order-37, first-order-38}. 

One can show that in certain class of theories the effective  potential calculated in one loop approximation, 
has indeed the desired form~\cite{temp-corr,temp-corr2}:
\begin{eqnarray}
V(\phi, T) = && \frac{1}{2} \left( m_0^2 + \alpha T^2 \right) \phi^2 +\frac{\lambda}{4} \ln \frac{\phi^2}{\sigma^2 \sqrt{e}}
\nonumber\\
&+& aT^4 \int_0^\infty  dx x^2\,\ln\left(\frac{1- e^{\sqrt{x^2 + b\phi^2/T^2}}}{1-e ^{-x}}\right) .
\label{eff-pot-2loop}
\end{eqnarray}
where $\sigma = 10^{14} - 10^{15} $ GeV, $\alpha$, $\lambda$, and $\beta$ are some  numbers which depend upon the 
coupling constant; the value of $a$ does not depend upon the interaction; the contribution of
the corresponding term at $\phi = 0$ gives simply thermal energy of particles. In the limit of low temperature we may
neglect the last term. This does not change qualitatively our conclusion.

Potential (\ref{eff-pot-2loop}) has a minimum at  $\phi = 0$ if $m_0^2 + \alpha T^2 > 0$. 
If $m_0^2 + \alpha T^2 < \lambda \sigma^2/e $, then there is another minimum at
$\phi \approx \sigma$. This minimum will be deeper than the first one and will
therefore be stable stable if $ m_0^2 + \alpha T^2 < \lambda \sigma^2 / 4$.
Thus the stable at thigh temperatures minimum of $V(\phi, T)$ at $\phi = 0$ transforms at lower  temperatures
into a quasistable one (the true vacuum of this theory)
separated from the original stable minimum by some  potential barrier.

As a rule, the probability of passage through the potential barrier is exponentially suppressed, 
so the system can remain in a quasi-stable state for a very long time. 
Tunneling in quantum theory was first considered in paper~\cite{VKO}. 
A more elegant method was proposed in~\cite{coleman-pt}
According to the results of these works, the probability of tunneling is determined by the value
of the action calculated on the solution of the classical equation of motion in imaginary
time. An approximate answer can be obtained in a simpler way by finding the
extremum of the action using the variational method. In particular, the
probability of tunneling per unit volume per unit time in potential (\ref{eff-pot-2loop}) at
zero temperature is equal to
\begin{equation}
\frac{dW}{dV dt} \approx M^4 \exp \left[- \frac{8\pi^2}{3 \lambda} \left(\ln \frac{\lambda \sigma^2}{m^2} \right)^{-1} \right],
\label{probability}
\end{equation}
where $M$ is some  unknown factor with dimension of mass.
Most likely, M is of the order of the inverse size of the bubble of the new phase,
i.e. $M \approx m$. Since $\lambda < 1$, the tunneling time is monstrously long and 
the universe will indeed have time to expand so much that no traces of matter, 
that was originally in it, would remain.  In reality, however, one should not think
that the temperature is equal to zero, because due to the presence
of the horizon, the temperature in the de Sitter universe cannot be lower
than $H/(2\pi)$~\cite{limT-DS} (in comoving coordinates).

Since the state with $\langle \phi \rangle = 0$ is unstable,
then, sooner or later the transition to the new phase with $\langle \phi \rangle \neq 0 $ should take place.
Immediately after formation of the bubble of the new phase, the magnitude of the classical field inside it
should be of the order $m / \sqrt{\lambda}$. This can be seen from the form of the dependence of the
effective effective potential (\ref{eff-pot-2loop}) on $\phi$. For natural values of parameters of the 
scalar field potential the value of $\phi$ quickly, in comparison with $H$, tends to its limiting value $\sigma$.
Indeed after the quantum jump creating the phase transition the evolution of $\phi$ is governed by the
classical equation of motion:
\begin{equation}
\ddot \phi + 3H \dot \phi = -\partial V / \partial \phi
\label{eqn-of-mot}
\end{equation}
with the initial conditions $\phi (0) = m /\sqrt{\lambda}$ and $\dot \phi (0) = 0$.
Here we neglected the term containing spatial derivatives since their contribution is divided  by the scale 
factor and quickly decreases with time. The value of $\partial V/\partial \phi$ at $\phi = m/\sqrt{\lambda}$ can be
estimated using Eq.~(\ref{eff-pot-2loop}) as $const \cdot m^2 \phi$, where is natural to expect that $m\gg H$.
Hence $\phi$ rises as $\exp(m t)$ and after the bubble formation the phase transition proceeds in time much shorter 
than $H^{-1}$ and thus an exponential  bloating  of the bubble does not take place.
In this case, there would be many bubbles in the visible today part of the universe, which in turn
would lead to too large density inhomegeneities.
More detailed discussion and references to the relevant literature can be found, for example, 
in ref~\cite{pt-first-order-37}.

In the modified version of the inflationary model~\cite{new-infl}  a
strong condition is imposed on potential $V(\phi)$, namely
$ m^2{T} \equiv \partial^2 V \partial \phi^2  \ll H^2$ which does not naturally follow from the theory.
This condition was specially invented to ensure a slow growth of the classical field
$ \phi(t)$ after the phase transition, $\phi(t) \sim \exp (m^2 t /3H)$. For a solution of standing 
in front of us problems it is sufficient that $m^2 <  H^2/25$.
In this case, the bubble's walls quickly disappear at infinity, and 
the entire universe is contained inside the interior of a single bubble,
the characteristic size of which at the moment of formation was of the order of $m^{-1}$ and which,
as a result of the universe expansion, reached the  size up to
\begin{equation}
r\gtrsim \frac{1}{r}\,\exp{(3H^2/m^2)} \cdot (T/(3K).
\label{r-larger}
\end{equation}
Here, T is the temperature of the primary plasma after the phase transition.
The total universe expansion in this version of  the model is by far larger and
differs from that given by eq.~(\ref{r-larger})
by the exponential factor presented in eq.~(\ref{probability}).
Naturally in this variant of the model, the inhomogeneities  will be significantly smaller, but
as shown by calculations in the frameworks of the standard $SU(5)$ theory, their magnitude 
happened to be larger than in reality approximately by two orders of magnitude~\cite{inhom-size}.

Let us note that in contrast to the previous case the inhomogeneities are not connected with the bubble walls,
but with rising quantum fluctuations in de Sitter universe. It is interesting that in the standard 
scenario of the universe evolution, i.e. without de Sitter stage, an estimate of inhomogeneities generated by
quantum fluctuations leads to the result approximately  two orders of magnitude smaller than it is necessary for
galaxy formation.

The condition of small inhomogeneities demands extremely slow variation of the effective 
potential $V(\phi)$ in rather large interval of variation of $\phi$, including the region of small
$\phi$. However, since the rate of  particle production by external field $\phi(t)$ 
is proportional to the speed of the field variation,
then in such a model, the particles that subsequently should fill the expanding void, forming
out world, are created too slowly and and their  density  and temperature, if they manage  to thermalize, 
turn out to be too small and it would be impossible to explain  the observed baryon
asymmetry of the universe.  Detailed discussion of these problem and the gravitational and 
temperature effects is presented in paper~\cite{first-order-38}. We note only that usually  the role
of gravity at the energies much smaller than the Planck one is not essential. However in the
model under consideration where a new hierarchy of masses is introduced, namely $m^2 \ll H^2$ or
$m^2 \ll R$, where $R$ is the four dimensional curvature of space-time. Just because of that the
role of gravitational corrections to the effective potential happens to be non-negligible. In particular
scalar field theory allows an addition to the Lagrangian the term $\xi \phi^2 R$ that essentially
changes the effective mass in de Sitter space,


The smallness of the effective mass $m^2$ in comparison with $H^2$ mentioned above 
implies smallness or cancellation of several terms each making a contribution to the coefficient 
in front of $ \phi^2/2 $ in the effective Lagrangian:
\begin{equation}
m^2 = m_0^2 + \alpha T^2 +\xi R + \lambda \langle \phi^2 \rangle + ...
\label{m-total}
\end{equation}
Here $m_0$ it the field mass in symmetric state in flat space-time  with zero temperature.
The natural value of $m_0$ in Grand Unification models is $10^{14}$  GeV.
An exception is presented by the Coleman-Weinberg model~\cite{col-win}  
which is actually defined by the condition $m_0 =0$, imposed on $V(\phi)$.
Perhaps there is some beauty in that, but strictly speaking, we do not have  any
basis for this assumption, as e.g. symmetry arguments, to impose  $m_0 = 0$.
In addition, it should be noted that the Coleman-Weinberg model has been
formulated in the flat space, where the condition $m_0 = 0$ means
\begin{equation}
\frac{\partial^2 V}{\partial^2 \phi}\large|_{\phi=0, R=0} = 0 ,
\label{m0=0-flat}
\end{equation}
where $R$ is the curvature of space-time. However, at $\phi =0$ the condition $R=0$ in inflationary model
is not fulfilled but instead of (\ref{m0=0-flat})
the relation $ R = - 8\pi G T_\mu^\mu = 32 \pi G V(\sigma)$, was suggested
in ref.~\cite{first-order-38} to change it to
\begin{equation}
\frac{\partial^2 V}{\partial^2 \phi}\large |_{\phi=0, R=32\pi GV(\sigma)}  = 0 ,
\label{m0=0-flat}
\end{equation}
Since we do not understand the origin of the cosmological term, the justification of the above condition looks
mysterious. In particular in the model of ref~\cite{dad-hidden-mass} the relation $R=32\pi GV(\sigma)$ 
generally speaking is not fulfilled. Hence the self-consistent formulation of the Coleman-Weinberg condition
should be modified.

The second term in eq.~(\ref{m-total}) arises due to interaction with thermal
bath in which field $\phi$ is situated. Then constant $\alpha$ is expected to be 
of the order of $N g^2$ where  $g \sim 0.5$ is the gauge coupling constant and
$n$  is the number of vector fields in the underlying symmetry group. Due to  existence
of the lower  limit on the temperature in De Sitter world $T_H = H/(2\pi) $~\cite{limT-DS} 
this term itself may break the necessary condition $m^2 < H^2 /25$.
 To prevent this from happening, in works~\cite{susy}
 where super-symmetric inflation was considered, it was assumed that $\phi$ is a gauge singlet 
and hence it does not interact with vector fields. Interactions with other fields may be made
arbitrarily weak and so constant $\alpha$ may be very small. This field which only role is to
 ensure inflation is called the inflaton. If we impose the condition of the conformal invariance
 at zero temperature we have to take $\xi = 1/6$. In this case the third term in eq.~(\ref{m-total}) gives too
 large contribution $\xi R = 12 \xi H^2 = 2H^2$. However we know the conformal invariance is, as a rule, broken,
 hence there is no necessity in this condition. In particular it can be shown that for Goldstone bosons
$\xi = 0$~\cite{xi-zero}.

The last term in eq. (\ref{m-total}) is induced by quantum fluctuations in curved space-time \cite{quant-fluc}.
It may be not essential if the self-interaction constant $\lambda$
of field $\phi$ is taken sufficiently small. So, leaving aside questions about the naturalness we see 
that it is possible to ensure 
sufficient duration of inflation after the phase transition. It is somewhat more difficult to solve
the problem of inhomogeneities  and generation of the baryon asymmetry. For that to be true  the
effective potential should be very smooth at $\phi <H$, so that $V'' (\phi) \ll H^2$ but very
abruptly falling down for a large $\phi$. Models leading to potentials of such a type exist but it
is still premature to say that the final version of the mechanism of exponential expansion is 
indeed found.

Let us briefly dwell on particle production in inflationary model. Immediately after phase transition the 
universe was formless and void, and darkness was upon it. There was no matter in the form of
elementary particles. The amplitude of $\phi$ rose in accordance with eq. (\ref{eqn-of-mot}).
The energy-momentum tensor in the r.h.s. of the equations of the General Relativity was given by the
expressions:
\begin{equation}
\rho= \rho_{vac} + V(\phi) + \dot\phi^2/2; \,\,\, p = - \rho_{vac} -V(\phi) + \dot\phi^2/2
\label{rho-p-of-phi}
\end{equation}
with $V(0) = 0$ and $V(\sigma)$ satisfying equation (\ref{average-T}), when $\phi$ reaches the value $\sigma$
which corresponds to  the stable minimum of the potential.

In the new inflationary model~\cite{new-infl} the tunneling, as it was already noted, goes to small $\phi=\phi_0$,
such that $ V(\phi_0) \ll \rho_{vac}$. It is also assumed  the field varies very slowly, so that $\dot \phi /\phi \ll, H$.
Hence at the first stage the character of expansion  practically would not be changed. Particle production at
this stage practically would not take place because of slow variation of $\phi$.  
Later when $\phi$ reaches sufficiently large values $V(\phi$ becomes steeper  and damped oscilations
of $\phi$ around equilibrium point $\sigma$ begin. Damping of oscillations is induced by the two reasons:
firstly by the universe expansion which is described by the friction term $3 H\dot \phi$ in eq. (\ref{eqn-of-mot})
and secondly by the particle production which is not explicitely taken into account in (\ref{eqn-of-mot}).
Since at large $\phi$ the oscillation frequency is quite high, $\omega = m (\phi=\sigma) = (10^{14}-10^{15})$ GeV, 
particle production at this period becomes quite essential. The expansion regime is drastically changed
going from the exponential (\ref{vacuum}) to the power law one (\ref{rel-regime}). Indeed for harmonic
oscillations the pressure given by eqs.~(\ref{rho-p-of-phi}) is zero. In our case deviations from harmonicity are not 
essential. The estimates made for several concrete models show that the rate of particle production
$\dot N /N$ (where $N$ is particle density in unit of space) is, as a rule, higher than the universe 
expansion rate,  $H = 2/(3t)$. It is assumed in the standard model that $\phi$ is the Higgs field, so its
couplings to other particles are proportional to the masses of the latter. Thus predominantly  heavist  
particles are created under condition that their muss is not too much larger then the oscillations 
frequency $\omega$. Hence the universe would be filled by superheavy boson in  strongly out of
equilibrium state. The last condition is favorable for generation of the baryon asymmetry of the universe.
After superheavy boson decays light particles such as leptons and quarks are created and the primeval
plasma was at last thermalized and acquires some temperature $T_1$ and the expansion law 
became relativistic (\ref{rel-regime}). The fact that the energy-momentum tensor was for a while dominated 
by heavy particles and the particle number density was smaller than the equilibrium one leads to
some diminishing of the baryon asymmetry after thermalization:
\begin{equation}
\beta = 3\beta_0 \,\frac{T_1}{m(\phi=\sigma)},
 \label{beta-fin}
\end{equation}
 where $m(\phi=\sigma)$ is the scalar field mass at the equilibrium point $\sigma$ and
$\beta_0$ is the baryon asymmetry originally created via decays of heavy bosons. Thus the
models in which the primeval plasma is cooled down too much to the moment of thermalization
 are excluded.

In addition to the difficulties described above, which are more technical than fundamental, the inflating universe 
theory faces a number of problems related to quantum tunneling in a gravitational field. The tunneling theory 
proposed in refs.~\cite{VKO,coleman-pt} for flat space was generalized to the case of 
tunneling in de Sitter space~\cite{tun-DS}.

However, the results of
these works cannot be directly applied to the case of interest, since the transition to the
imaginary time, which lies at the heart of the method used, was performed for the exact
de Sitter space, while the real universe is not such, and 
deviations of the exact solution from the approximate one may be significant.

Here it might make sense\footnote{ Analogous arguments have been used by A. Goncharov and A. Linde,
private communication.}
to use the Hamilton formalism and to try to solve the
functional Schrödinger equation in quasi classical approximation (there is no other 
known way anyhow), that describes the considered quantum field theory in the external  curved
metric.  The neglect of the back reaction of the field on gravity is justified by fact that the condition
of a long inflationary phase after formation of the bubble of the new phase is equivalent, as one
can easily see, to the demand that the vacuum energy changes very little after the quantum jump.

Assuming the the metric has the form $ds^2 = dt^2 - e^{2Ht} dr^2 $ we obtain for the wave functional
$\Psi (\phi)$ the the following equation:
\begin{equation}
 \left[ 2i \frac{\partial}{\partial \tau}+ \frac{\delta^2}{\delta \phi^2} - 
 \frac{2}{9 H^2 \tau^2} \int d^3 x [\,\frac{1}{2} (3H\tau)^{2/3} (\nabla \phi)^2 +V(\phi) ] \right]\Psi = 0,
\label{delta-Psi}
\end{equation}
where $\tau = e^{-3Ht}/(3H) $,  and $V(\phi)$ is the effective potential of field $\phi$:
 \begin{equation}
 V(\phi) = \frac{1}{2} m^2 \phi^2 + \frac{\lambda}{4} \phi^4\,\ln\frac{\phi^2}{\sigma^2} .
 \label{V-of-phi2}
\end{equation} 
Thus there arises a problem of tunneling in time dependent potential.

An example of significantly simpler one dimensional (and not infinitely dimensional) 
quantum mechanical problem as (\ref{delta-Psi}) with the potential $U = \tau^{-n} v (x)$
shows that for $\tau \rightarrow  0$ the usual expression for the tunneling probability
\begin{equation}
\Gamma \sim \exp \left[-\int dx \sqrt{2mU} \right]
\label{Gamma-1dim}
\end{equation}
is applicable if $n>2$ but is not if $n<2$. In the case $n=2$ the result depends upon the 
parameter of the potential $V(\phi)$ and eq.~(\ref{Gamma-1dim}) is valid if the coefficient in front of 
$x^2$ is sufficiently large.  If this result is directly applied to eq.~(\ref{delta-Psi}), then one can see 
that for the validity of the improved inflationary model the situation is opposite, $m^2\ll H^2$. But in
this case quasiclassical approximation is not applicable. Thus finally we do not have an adequate 
formalism for description of tunneling in gravitational field.
 
For  realization of inflationary model the value $\phi_0$, which takes field $\phi$ after tunneling, is
of primary importance. For potential (\ref{V-of-phi2}) in flat space-time the  value $\phi_0 = m/\sqrt{\lambda}$
is sufficiently small to lead to slow motion of $\phi$ to the limiting value $\sigma$, not destroying
long exponential expansion. On the opposite if $\phi_0$, is large, the equilibrium state is reached 
quickly and exponential expansion turns into the power law one. One  can show that if the size
of the created bubble of the new phase  is $r$, the magnitude of the field in this bubble is
 $(\sqrt{\lambda}r)^{-1}$. Thus for successful inflation large bubbles are necessary. 

However in the theory described by the non-stationary equation (\ref{delta-Psi}), 
the bubble size is unknown. In the case considered in ref~\cite{tun-DS} it is shown that the bubble 
size does not exceed $H^{-1}$ that is natural, because this is the horizon size in De Sitter space.
Still it is unclear if this result is a consequence of the thin wall approximation used in the quoted papers
or, which is probably more important, that the universe is not exactly the De Sitter one. The point is  that
the transition to imaginary time leads to the transformation of the De Sitter space into four dimensional
sphere of radius  $H^{-1}$ so the size of the bubble in three dimensional space in the moment of its
formation cannot exceed  that. If this result survives in the real situation, the inflationary model could be
in serious danger, because in this case  $\phi_0$ would be big and hence a large expansion of the bubble is 
impossible.

Another point could be serious is that we use effective Lagrangian assuming that the fields are slowly
changing but their variation  in the expanding world is not so small, generally speaking it is
$\dot \phi/\phi \approx H$. So we need to take into  account loop corrections (but how?) not assuming 
$\phi = const$. It is not clear how all that may influence  on the tunneling and the value of $\phi_0$.

The super-small size of the region from which our Universe began to inflate is also often a source of 
concern. According to equations  (\ref{r-larger}) and  (\ref{probability}) even for a rather modest value
$\lambda =0.1$ the size of the region which now makes all the visible universe was surely smaller
than, say,  $10^{-100} $ cm. It is difficult to agree that at so small distances, even in vacuum
(which as we now know is quite complicated) a serious modifications of the known to us
physical laws have not took place. In my opinion there is no reasons for anxiety, because one  can always
speak about exponential expansion of a sufficiently large regions where no surprise in the vacuum structure  
happens. Even if in the course of inflation of very small regions some unknown phenomena appear,
they should disappear when the size of these regions becomes sufficiently large,

Recently an interesting version of inflationary model for which no phase transition is necessary
was suggested by Linde~\cite{linde-aa}. The starting point of this model is the assumption that
at some initial moment scalar field $\phi$ might take very large values 
$\phi \gg m_{Pl} \approx 10^{19} $ GeV,  with spatial variation of $\phi$ being sufficiently low.
Such a situation can be realized in the case of chaotic initial conditions if the self-interaction
coupling constant is small, so that $V = \lambda \phi^4/4 < m_{Pl}^4$. In this case at the right 
hand side of evolutionary equation (\ref{eqn-of-mot}) the term $H \dot \phi$ starts to dominate,
where $H \eqsim (8\pi \rho m_{Pl}^{-2} /3)^{1/2} \approx (2\pi\lambda /3)^{1/2} \phi^2 m_{Pl}^{-1}$
and the solution of this equation takes the form
\begin{equation}
\phi= \phi_i \exp\left[ -\frac{\sqrt{\lambda}}{\sqrt{6\pi}}\, m_{Pl} t \right] .
\label{phi-sol2}
\end{equation}
Hence the expansion rate of the universe $\dot a/a =H$ happens to be larger than the rate
of $\phi$ decrease due to the factor $\phi_i /m_{Pl} \gg 1$ and the universe region, where 
the conditions mentioned above were accidentally, has expanded exponentially:
\begin{equation}
\frac{a}{a_0} = \exp \left( 2\pi\,\frac{\phi_i^2}{m_{Pl}^2} \right) . 
\label{inflating-a}
\end{equation}

This expansion could provide a solution of the problems discussed above if the ratio
$\phi_i^2/m_{Pl}^2$ is sufficiently large, namely $\phi_i^2/m_{Pl}^2 \gtrsim 10$. So with
chaotic initial conditions in infinite universe there always could found a region that strongly 
exponentially expanded and as a result reach the state suitable for our existence.
Other uncomfortable regions of the universe would be outside of possibilities of our observations

For realization of this model it is not necessary to impose many special condition on the field theory, 
it is enough to use a simple hypothesis on existence of weakly interacting  and self-interacting
field $\phi$, i.e the assumption of a small $\lambda$ and weak coupling to other fields.
Not yet worked out is the problem of of quantum gravity corrections to the classical equations of
of motion for a large $\phi$.

Of course there is still a question about naturalness of the initial conditions.
In contrast to the classical Friedman cosmology, where a very precise 
fine tuning of the initial state is demanded, here we have the stochatsically distributed field
$\phi$ in chaotic universe near singularity.  In my opinion the hypothesis about initial chaos
is much more attractive and this variant possibly corresponds to the real case, though
the question about the origin of the initial chaotic state remains open.\\[2mm]
{\bf  Conclusion}\\[1mm]
So  presently there are two principally different approaches to the problem of the origin and
evolution of the universe. The first one, to one or other degree, is based on the anthropic principle, according to 
which the fact that life existence in the universe makes senseless the question why the universe is such
but not other. This approach cannot be denied the right to exist, especially if there are an infinite set
of different universes  is realized. Then out of this chaotic set only a few universes 
with very specific conditions could be available for us. However from  the point of view of the anthropic
principle the colossal  redundancy of other galaxies is mysterious.

In other approach is assumed that the universe is one and only one but initial conditions there is 
arbitrary. However, and this lies in the basement of all theoretical models, the laws of physics are
such that practically from any initial state we arrive to our very non-trivial world.

Inflationary model discussed above responds positively  to the latter demand. However, the version
with chaotic universe is an intermediate one between those two approaches. It goes without saying 
that this model cannot  be considered as the final theory. From one side there are some unsolved problems
inside the model, such as e.g. that concerning tunneling in the expanding world.
On the other hand it is not established on which field theory this model is based. It s difficult to expect 
to find the solution to the last problem until the elementary particle theory is not worked out 
that is applicable up to the Planck energies. Sooner it is another way around, if it is assumed that
inflationary model based on strongly delayed  first order phase transition, does indeed correctly describes
reality, one could derive conditions that the elementary particle theory must satisfy. 

What ground we have to believe that the model of inflationary universe is really true?
First, it is beautiful, since it is based on one very simple condition on existence of 
De Sitter stage some time in the past. This allows to solve  in a uniform way the problems
of homogeneity, isotropy, horizon, flatness, and relic magnetic monopoles. These facts are
surely in favor of this scenario. Against the the model, though indirectly, is the problem of the
cosmological constant and to a smaller (up to the present time) extent and an absence of the detailed 
theoretical scheme. The status of inflationary model would be very much stronger if it
confirmed that the cosmological parameter $\Omega = \rho/\rho_c$ is equal to 1.
Unfortunately at the present time it is not seen if this can be established with sufficient accuracy
On the opposite it seems it is easier to reject the model obtaining an upper bound on $\Omega$.
 
Still even if inflationary model obtains convincing evidence in its favor (most probably they will be
theoretical but not observational )
still complete happiness will be far away until the 
approaches to solving the two remaining critical problems have been found:
the problem of cosmological constant and the problem of the universe creation and singularity.
However, we mustn't forget that "appetite comes with eating" -
after all, quite recently, those fundamental problems that we now, thanks to the inflationary model, 
consider already solved, or we say (those who are more cautious) that there appears a possibility
of their solution. A few years ago, they seemed completely impregnable,
and the importance of this achievement should not be underestimated.

\section*{Appendix - translation of the  content of Einshteinovskij Sbornik
}
 A. Einstein. How the Theory of Relativity Was Created.\\
B.E. Yavelov, V. Ya. Frenkel. On Some Historical and Physical Aspects of the Einstein-de
Haas Experiments.\\
V.Ya. Frenkel, B.E. Yavelov. "This is What Can Happen to a Person Who Thinks a Lot
but Reads Little."\\
B.G. Kuznetsov. The Einstein-Bohr Collision, the Einstein-Bergson Collision, and Science
in the Second Half of the Twentieth Century.\\
V.P. Vizgin. Einstein, Hilbert, Weyl: The Genesis of the Program of Unified Geometrized
Field Theories.\\
A. Salam. Einstein’s Final Vision: Unifying Fundamental Interactions and the Properties
of Space-Time.\\
A.D. Dolgov. Progress in Particle Physics and Modern Cosmology.\\
B. M. Bolotovsky. On the Apparent Form of Rapidly Moving Bodies.\\
G. A. Lorentz. On Einstein’s Theory of Gravitation.\\
T. Levi-Civita. On an Analytical Expression for the Gravitational Tensor in Einstein’s Theory.\\
E. Schr{$\ddot o$}dinger. Components of the Gravitational Field Energy.\\
G. Bauer. On the Components of the Gravitational Field Energy.\\
G. Nordstr{$\ddot o$}m. On the Gravitational Field Energy in Einstein’s Theory.\\
F. Klein. On the Integral Form of the Conservation Laws of the Theory of a Spatially
Closed World.\\
L. Rosenfeld. On the Gravitational Actions of Light.\\
M.P. Bronstein. Quantum Theory of Weak Gravitational Fields.\\
M.P. Bronstein. On the Possibility of Spontaneous Photon Splitting.\\
G.E. Gorelik, V.Ya. Frenkel. M.P. Bronstein and His Role in the Development of the
Quantum Theory of Gravity.\\
I.Yu. Kobzarev. Review of W. Rindler’s book "Foundations of the Theory of Relativity
(WHAT, GTR, and Cosmology)"

\end{document}